\documentclass[
 reprint,
 superscriptaddress,
 amsmath,amssymb,
 aps,
pra,
]{revtex4-2}
\usepackage{subfig}
\usepackage{times,txfonts}
\usepackage{braket}
\usepackage{quantikz}
\usepackage{graphicx}
\usepackage{dcolumn}
\usepackage{bm}
\usepackage[colorlinks,linkcolor=blue,urlcolor=blue, citecolor=blue]{hyperref}

\newcommand{\ketbra}[2]{|#1\rangle\! \langle #2|}

\def\captionof#1#2{{\def\@captype{#1}#2}}

\DeclareMathOperator\arctanh{arctanh}

\begin{document}

\preprint{APS/123-QED}

\title{DQC1 as an Open Quantum System}

\author{Jake Xuereb}
\email{jqed.xuereb@gmail.com}
\affiliation{Department of Physics, University of Malta, Msida MSD 2080, Malta}
\affiliation{School of Physics, University College Dublin, Belfield, Dublin 4, Ireland}
\affiliation{School of Physics, Trinity College Dublin, College Green, Dublin 2, Ireland}

\author{Steve Campbell}
\affiliation{School of Physics, University College Dublin, Belfield, Dublin 4, Ireland}
\affiliation{Centre for Quantum Engineering, Science, and Technology, University College Dublin, Belfield, Dublin 4, Ireland}

\author{John Goold}
\affiliation{School of Physics, Trinity College Dublin, College Green, Dublin 2, Ireland}

\author{Andr\'e Xuereb}
\affiliation{Department of Physics, University of Malta, Msida MSD 2080, Malta}

\begin{abstract}
The \textbf{DQC1} complexity class, or power of one qubit model, is examined as an open quantum system. We study the dynamics of a register of qubits carrying out a \textbf{DQC1} algorithm and show that, for any algorithm in the complexity class, the evolution of the logical qubit can be described as an open quantum system undergoing a dynamics which is unital. Unital quantum channels respect the Tasaki-Crooks fluctuation theorem and we demonstrate how this is captured by the thermodynamics of the logical qubit. As an application, we investigate the equilibrium and non-equilibrium thermodynamics of the \textbf{DQC1} trace estimation algorithm. We show that different computational inputs, i.e. different traces being estimated, lead to different energetic exchanges across the register of qubits and that the temperature of the logical qubit impacts the magnitude of fluctuations experienced and quality of the algorithm.
\end{abstract}

	\maketitle

\maketitle

\begin{figure*}[t]
\centering
\includegraphics[width = \textwidth]{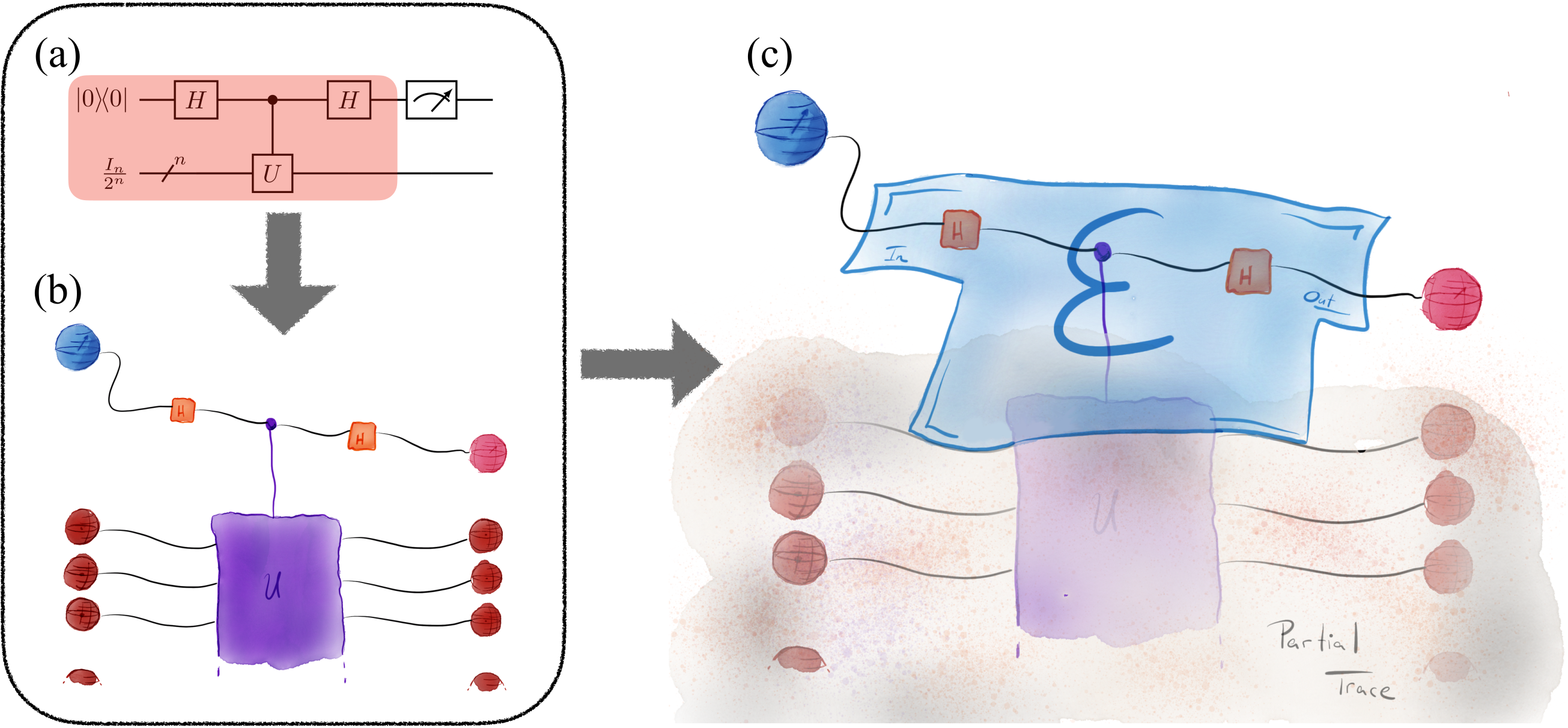}
\caption{The main ideas of this work visualised, (a) a quantum circuit can be thought of as (b) an effective open quantum system whose dynamics be analysed through the formalism of (c) quantum channels.}
\label{fig1}
\end{figure*}

\section{Introduction}
A quantum algorithm in gate based quantum computation is a process where the information content of a quantum state is changed towards achieving a computational goal upon measurement. Since information processing and thermodynamics can, in certain contexts, be two sides of the same coin one would expect that relationships can be made between a computational task and the thermodynamics experienced by the system carrying out an algorithm to achieve this goal.



It is well established that the ability to manipulate information is intimately related to the physical architectures upon which it is encoded. However, despite large strides having been made towards understanding the relationship between quantum information and thermodynamics \cite{Palma,Horodecki_2013,Brand_o_2015,Huber_2015,goold_modi_heat_exchange,goold_modi_landauer,taranto2021landauer,Faist_2019}, the thermodynamics of quantum computation itself, as is highlighted in~\cite{auffeves2021quantum}, has been largely unexamined despite its clear importance. Recent work  has started to move in this direction with the study of thermodynamics of quantum annealers~\cite{CampisiQST,CampisiPRE,DeffnerSciRep}, investigations of heat engines in quantum computation~\cite{CampisiPRXQ,Landi_2020} and thermodynamic optimisation of information processing tasks focusing on changes in numbers of qubits~\cite{meier2021thermodynamic,taranto2021landauer}. 

In this work we explore a complementary theme by focussing on the thermodynamics of a quantum system in light of the content of the computation it is carrying out. In particular, we consider arguably the simplest quantum complexity class as captured by the deterministic quantum computation with one clean qubit model, DQC1. Beyond its relative simplicity, this class of algorithms is particularly informative to explore as it is readily implementable in experimental architectures such as NMR setups. We examine the computation using the formalism of quantum channels, allowing to model the process as an effective open quantum system. This picture then allows to explore the energetics and thermodynamics at play. We demonstrate our results explicitly for the trace estimation algorithm, establishing that the magnitude of the trace to be determined directly impacts the energetic exchanges and fluctuations.

The remainder of the work is organised as follows. After introducing the \textbf{DQC1} complexity class in Section~\ref{sec:classreview}, we prove our main result in Section~\ref{sec:unitality_dqc1} by showing that the dynamics of the logical qubit of any \textbf{DQC1} algorithm is unital \cite{Mendl_2009,watrous_2018} in the sense of open quantum systems. In Section~\ref{sec:Crooks} we discuss how this implies that these dynamics satisfy the Tasaki-Crooks fluctuation theorem \cite{crooks,tasaki,Talkner_2007,Campisi_2011}
and in Section~\ref{sec:dqc1_thermo} present the thermodynamics of the trace estimation algorithm as an application. Here we find that different trace estimations lead to different energetic exchanges amongst the qubits of the \textbf{DQC1} register and that the temperature of the control qubit in this algorithm impacts the amount of fluctuations one expects in this model. 

\section{The \textbf{DQC1} Complexity Class}
\label{sec:classreview}
The deterministic quantum computation with \textit{one-clean-qubit} model (\textbf{DQC1}) is a restricted class of quantum algorithms satisfying the following properties:
	\begin{enumerate}
		\item the initial state is $\rho = \ketbra{0}{0} \otimes I_n/2^n$, that is a \textit{clean} qubit in a pure state and an $n$-qubit ancillary register in the maximally mixed state;
		\item a poly($n$)-size gate can be enacted across the $n+1$ qubits;
		\item the \textit{clean} qubit is measured giving the output bit $a \in \{0,1\}$.
	\end{enumerate}
Introduced by Knill \& Laflamme to investigate the \textit{power of one bit of quantum information}~\cite{DQC1}, \textbf{DQC1} was inspired by nuclear magnetic resonance qubits (NMR)~\cite{laflamme2002introduction} where experimentally a global thermal state of such nuclear spins is simple to achieve by equilibration and a single spin can be addressed changing its magnetic polarisation relative to the bath giving register of qubits capable of \textbf{DQC1} algorithms in its idealised limit. Since then, \textbf{DQC1} has also been implemented across different architectures such as trapped ions~\cite{zhang_2019} and photonic setups~\cite{Lanyon_2008,Hor_Meyll_2015}. Away from experiment, it has been studied as a complexity class in its own right~\cite{shepherd2006computation,morimaei_2014,Fujii_2018,aaronson2016computational}. It is capable of solving various problems, \cite{integrability,shepherd2006computation,shor2008estimating,cade2017quantum} some of which are classically hard and has also been studied as a potential quantum machine learning model~\cite{dqc1_qml}. In particular, the trace estimation of unitary matrices has been shown to be \textbf{DQC1}-complete~\cite{shepherd2006computation}. All this despite being incapable of universal quantum computation~\cite{ambainis2000computing}.

The initial state of \textbf{DQC1} circuits may be expressed as 
	\begin{gather}
	\rho = \ketbra{0}{0}\otimes\frac{I_n}{2^n} = \frac{ I_1 \otimes I_n +\sigma^z\otimes I_n}{2^{n+1}} = \frac{ I_{n+1} +\sigma^z_1 }{2^{n+1}},
	\end{gather}
	where $\sigma^z$ acts on the first qubit. This means that the state at the end is given by 
	\begin{gather}
	\rho' = \frac{ I_{n+1} +V\sigma^z_1V^\dagger}{2^{n+1}},
	\end{gather}
	where $V$ is the product of unitaries implemented across the \textbf{DQC1} register during the computation.
	If one then defines~\cite{shepherd2006computation}
	\begin{align}
		\mu = \frac{\text{tr}\{V\sigma^z_1V^\dagger\sigma^z_1\}}{2^{n+1}} && R = \frac{V \sigma^z_1V^\dagger - \mu \sigma^z_1}{\sqrt{1 - \mu^2}}, \nonumber
	\end{align}
	the final state before measurement can be expressed as
	\begin{gather}
	\rho' = \frac{I_{n+1} + \mu \sigma^z_1 + \sqrt{1-\mu^2}R}{2^{n+1}},
	\end{gather}
	where $R$ and $\sigma^z_1$ are traceless and $\mu$ is real. The probability of measuring $0$ on the first qubit is then 
	\begin{gather}
	\mathcal{P}[0] = \text{tr}\left\{\rho'\frac{I_1+\sigma^z_1}{2}\right\} = \frac{1+\mu}{2}. \nonumber
	\end{gather}
	Applying this structure more specifically to estimate the trace of some $U$ using the circuit in Figure~\ref{fig1}~(a) we have 
	\begin{gather}
	\rho' = \frac{I_{n+1} + \sigma^z_1\otimes\left(\frac{U + U^\dagger}{2}\right) + \sigma^y_1\otimes\left(\frac{U-U^\dagger}{2i}\right)}{2^{n+1}},
	\end{gather}
	and therefore measuring with respect to $\sigma^z_1$ gives $\mu =\frac{\text{Re} \{\text{tr}U\}}{2^{n}}$.

The work of Linden \& Jozsa~\cite{Jozsa_2003} demonstrated that quantum computation involving pure states necessarily requires the establishment of multipartite entanglement. However, the situation is less clear cut when dealing with mixed states as in \textbf{DQC1}, particularly when we allow for the logical qubit to also be in a mixed state. We can consider a generalisation of the \textbf{DQC1} trace estimation algorithm when the first qubit is not \textit{clean} but possesses some polarisation $\alpha$ 
\begin{gather}
\rho(\alpha) = \frac{I_1+\alpha\sigma^z}{2}\otimes \frac{I_n}{2^n} \nonumber\\
\rho'(\alpha) = \frac{I_{n+1}+ \alpha\sigma^z_1\otimes\left(\frac{U + U^\dagger}{2}\right) +\alpha\sigma^y_1\otimes\left(\frac{U-U^\dagger}{2i}\right)}{2^{n+1}}, \nonumber
\end{gather}
as introduced in~\cite{datta_2005} as a prototypical circuit to study the resource giving advantage is this setting. Here it has been shown that no distillable entanglement, quantified by negativity, is formed across the first qubit and the fully-mixed qubits but that non-negligible amounts of negativity are found in bipartitions of the fully mixed qubits and the first qubit. In later work~\cite{datta_2008}, a measure of non-classical correlations called the \textit{quantum discord} \cite{discord_01,discord_01_2} was shown to be non-zero between the first qubit and the mixed qubits leading to the argument that the quantum discord is the resource behind the speed-up in trace estimation offered by \textbf{DQC1}. This claim has been argued against in~\cite{dispute_2010} by counterexample and more recently in a very robust argument made by Cade and Yoganathan who showed that the \textbf{DQC1} trace estimation circuit is classically simulable for the class of unitaries that do not generate entanglement, that is \textbf{DQC1}$_{sep}\subseteq$ \textbf{BPP} \cite{cade2017quantum}.

This generalisation is referred to as the \textit{one-non-clean-qubit} model and was studied in~\cite{non_clean} as a computational model in its own right.  This model is more representative of what is physically achievable within an NMR architecture since the \textit{clean} qubit will always have some degree of polarisation. As such, knowing how this polarisation imparts error in the computation is relevant.

\section{Unitality of \textbf{DQC1} Channels}
\label{sec:unitality_dqc1}
Having introduced the \textbf{DQC1} complexity class, we now demonstrate that the dynamics of the logical can be recast as a quantum channel using the tools of open quantum systems, illustrated in Fig~\ref{fig1}~(b,c).

In this formalism we may associate this evolution with a Stinespring dilated representation as 
\begin{gather}
    \mathcal{E}(\rho_S) = \text{tr}_n\left\{V\left(\rho_S \otimes \frac{I_n}{2}\right)V^\dagger\right\} = \rho'_S,
\label{DQC1channel}
\end{gather}
where $V$ is a unitary satisfying property 2 representing the product of unitary gates carried out on the register which result in a specific \textbf{DQC1} algorithm.
By the Choi-Kraus Theorem we can associate a Kraus representation to this map
\begin{gather}
  \mathcal{E}(\rho_S) = \sum_{i}\Lambda_i \rho_S \Lambda^\dagger_i \, : \, \sum_{i} \Lambda_i^\dagger \Lambda_i = I_1,
\end{gather}
where $\Lambda_i$ are the corresponding Kraus operators which are derived from the unitary $V$ in the Stinespring dilation as \begin{gather}
    \Lambda_{i,j} = \sqrt{b_j}\bra{i_B}V\ket{j_B},
\end{gather}
where $\ket{i_B}$, $\ket{j_B}$ are eigenkets of the environment and $b_j$ are the eigenvalues of the initial state of the environment. Note that we are using two indices since the environment is in a mixed state. In the \textbf{DQC1} setting we assume the ancillary register is in the maximally mixed state and plays the role of the environment. The state of our environment is diagonal in the energy eigenbasis of a local Hamiltonian $H= \sum_i E_i\ketbra{E_i}{E_i}$ since it can be expressed $\frac{I_n}{2^n}= \frac{1}{2^n} \sum_i\ketbra{E_i}{E_i}$.
As such, the Kraus operators $\Lambda_i$ can be expressed directly as
\begin{gather}
    \Lambda_{i,j} = \frac{1}{\sqrt{2^n}}\bra{E_i}V\ket{E_j},
\end{gather}
and considering 
\begin{align}
\sum^{2^n}_{i,j=1} \Lambda_{i,j}\Lambda^\dagger_{i,j} &=\sum^{2^n}_{i,j=1}\frac{1}{2^n}  \bra{E_i}V\ketbra{E_j}{E_j}V^\dagger\ket{E_i} \nonumber\\
&= I_1, \nonumber
\end{align}
by completeness of the energy eigenbasis and unitarity of $V$, we have that $\mathcal{E}\left(\cdot\right)$ is a unital channel for any \textbf{DQC1} algorithm. This is our first result. 

To understand the origin of unitality we can consider instead the register to be in an arbitrary mixed state giving
\begin{gather}
\sum^{2^n}_{i,j=1}\Lambda_{i,j}\Lambda^\dagger_{i,j} =  \sum^{2^n}_{i,j=1} \sqrt{b_j}\bra{E_i}V\ketbra{E_j}{E_j}V^\dagger\ket{E_i},
\end{gather}
where now we are unable to use the completeness relation unless $b_j$ is either constant, as in the case of a maximally mixed state, or has only a single non-zero contribution, as in the case of the ancillary qubits being in the ground state. The unitality of \textbf{DQC1} channels thus emerges as a result of the environment, i.e. the ancillary register found in the maximally mixed state and property 1 of this complexity class. We explore unitality for more general quantum circuits in Appendix~\ref{sec:Unitality}, in particular communicating that any circuit involving separable unitaries across a register of qubits, such as local rotations - can be described by a unital channel. In this sense unital channels capture the dynamics of a relevant component of quantum computation. This further investigation shows that the dynamics of non-separable i.e. entangling unitaries is more complex since the unitality of the dynamics is dependent both on the state of the environment, and so the choice of bipartition across the register, and also the unitary carried out.

To summarise, \textbf{DQC1} algorithms encode all relevant computational information in the state of the logical qubit and this is the only qubit which is measured as described by property 3. In light of this, it is relevant to examine the evolution of this qubit throughout this quantum information process. We have shown that the dynamics of this evolution is unital. In the following sections, we will apply this fact to the examine the thermodynamics and energetics of this complexity class and specifically of the trace estimation algorithm.

\section{\textbf{DQC1} \& the Tasaki-Crooks Fluctuation Theorem}
\label{sec:Crooks}
Quantum channels \cite{nielsen_chuang_2010,watrous_2018,modi_operational_dynamics} provide an operational formalism to describe open quantum system dynamics without the direct use of a master equation~\cite{petruccione_breuer}. As such, they carry information about the thermodynamics of the process they describe. In particular, it has been shown that unital channels~\cite{Mendl_2009,watrous_2018} model processes that satisfy the Tasaki-Crooks fluctuation theorem \cite{albash_2013,rastegin_2013,goold_modi_AVS} which relates the ratio of the forward and reverse work probability distributions to the entropy produced by the process. Therefore, as a corollary of the result from Section~\ref{sec:unitality_dqc1}, we have that the dynamics of the logical qubit in any \textbf{DQC1} algorithm necessarily respects the Tasaki-Crooks Theorem. In this section we explore this corollary and its implications in the context of \textbf{DQC1} trace estimation with a non-clean qubit as examined in \cite{datta_2005,datta_2008,yoganathan2019clean} to gain an operational understanding of what the Fluctuation theorem could mean for a quantum algorithm.

Consider a \textbf{DQC1}-channel Eq.~\eqref{DQC1channel} with $V = (H \otimes I_n)\left(\ketbra{0}{0}\otimes I_n + \ketbra{1}{1}\otimes U\right) (H \otimes I_n)$ where $H$ is the Hadamard gate on the logical qubit, $I_n$ is the identity operation on the $n$-qubit ancillary register and $U$ is the unitary which is applied to the ancillary register conditioned on the logical qubit. It is the trace of this $U$ which we estimate by measuring the logical qubit.

The logical qubit in this algorithm is treated as the energetic system of interest, with initial state 
\begin{gather}
\rho_C(\alpha) = \frac{I_1+ \alpha \sigma^z_1}{2},
\label{rhoinit}
\end{gather}
where $I_1$ is the single qubit identity matrix and $\alpha$ is a coefficient determining its degree of polarization or mixedness. This state is a locally thermal state with respect to the system Hamiltonian
\begin{gather}
H_C = -\omega \sigma^z \, : \, \beta = \omega^{-1} \arctanh (\alpha),
\label{eq:initial_temp}
\end{gather}
and being a single qubit state we can associate an effective inverse temperature $\beta$ to it where $H_C$ is a typical qubit Hamiltonian with $\omega$ a constant related to the Larmor precession frequency on the Bloch Sphere. The final state of the logical qubit in the trace estimation protocol before measurement is 
\begin{gather}
\rho'_C(\alpha) = \frac{I_1 + \alpha\frac{\text{Re}\{\text{tr}U\}}{2^n}\sigma_1^z}{2},
\label{rhofinal}
\end{gather}
for the trace estimation of unitaries with only real contributions. This constraint on the computational input ensure that the state is always locally thermal with respect to $H_C$.

\begin{figure}[t!]
\centering
\includegraphics[width = 0.45\textwidth]{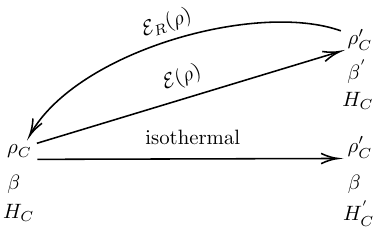}
\caption{The setup for our operational analysis as discussed in~\cite{goold_modi_AVS,deffner_2011}.}
\label{fig2}
\end{figure}
The Tasaki-Crooks Fluctuation theorem for the logical qubit in this algorithm can now be investigated where we consider the reversed channel $\mathcal{E}_R(\cdot)$, derived below, which models the reverse process the system can fluctuate to. Using a two-point-projective-measurement scheme (TPM) \cite{work_not_an_observable} on the setup presented in Figure~\ref{fig2}, work distributions for the forward and reverse processes are obtained allowing us to derive the Fluctuation Theorem in this context. In Figure~\ref{fig2} we also depict an isothermal process as is typical of setups for this derivation \cite{goold_modi_AVS,deffner_2011}, carried out by an external agent with local control over the logical qubit who changes the Hamiltonian's precession frequency to 
\begin{gather}
    \omega' = \omega\left(\arctanh(\alpha)\right)^{-1}\arctanh\left(\alpha\frac{\text{Re}\{\text{tr} U\}}{2^n}\right).
\end{gather} For a more extensive discussion on the association of an effective temperature to these states refer to Appendix~\ref{sec:effective_temp}.

It will prove useful to convert this channel to its Choi representation (see Appendix~\ref{Appendix:Choi} for a detailed derivation)~\cite{modi_operational_dynamics,choi,jamiolkowski} 
\begin{gather}
\Upsilon_\mathcal{E} = \left(\mathcal{E}(\rho) \otimes I_{1}\right)\ketbra{\varphi}{\varphi},
\label{choi1}
\end{gather}
where $\ket{\varphi} = \sum^{2}_{i = 1} \ket{ii}$ i.e. the two qubit un-normalised Bell state.

The impact of a channel on a quantum state making use of its Choi representation is given by $    \mathcal{E}\left(\rho_S\right) =  \text{tr}_{B}\left\{\left({I_S}\otimes\rho_S^\text{T}\right)\Upsilon_\mathcal{E}\right\}$ so
it is easy to verify that
\begin{gather}
    \mathcal{E}\left(I_{1}\right) =  \text{tr}_{2}\left\{\left(I_{1}\otimes {I_1}^\text{T}\right)\Upsilon_\mathcal{E}\right\} = I_{1}
\end{gather}
where $\text{tr}_{2}\{\cdot\}$ refers to tracing over the second d.o.f, meaning that the identity is a fixed point of this map and that this Choi state represents unital dynamics, corroborating the result of Section~\ref{sec:unitality_dqc1}. We now define the time-reversed channel for this process as 
\begin{gather}
    \mathcal{E}_R(\rho) = \text{tr}_n\left\{V^\dagger\left(\rho \otimes \frac{I_n}{2}\right)V\right\}
\end{gather}
which has a Choi state that is the complex conjugate of Eq.~\eqref{choi1} and as such is also unital giving 
\begin{gather}
    \frac{\mathcal{P}_F(\Delta E_C)}{\mathcal{P}_R(\Delta E_C)} = e^{\beta\left(\Delta E_C - \Delta F\right)}
\end{gather}
where $\mathcal{P}(\Delta E_C)$ is an energy distribution obtained through the TPM across the forward or backwards channel, as indicated by the subscript, and $\Delta F$ is the free energy change in the isothermal process, following the presentation of~\cite{goold_modi_AVS}.

As the overall dynamics is unitary, we can analyse the energetic exchanges between the logical qubit and the register. As shown in Appendix~\ref{sec:energetic_laws}, since the register remains in a maximally mixed state throughout the process, there is no heat exchange from the logical qubit to the environment and therefore
\begin{equation}
    \Delta E_c = \langle W_C \rangle.
    \label{control_qubit_1stA}
\end{equation}
We find that the work done, $\langle W_C \rangle$, on the logical qubit is solely due to the heat from the register in the trace estimation algorithm giving
\begin{gather}
    \frac{\mathcal{P}_F(W_C)}{\mathcal{P}_R(W_C)} = e^{\beta\left(W_C - \Delta F\right)} =e^{\Sigma_C}
\end{gather}
where $\Sigma_C$ is the entropy production of the logical qubit discussed in Appendix~\ref{sec:energetic_laws}. By using the generalised Landauer equality~\cite{esposito_eq,landauer_2014} in this setting, cfr. Sec.~\ref{sec:energetic_laws}, we obtain the Tasaki-Crooks Fluctuation theorem for the logical qubit
\begin{gather}
    \frac{\mathcal{P}_F(W_C)}{\mathcal{P}_R(W_C)} = e^{\Delta S_C}.
\label{crooks}
\end{gather}
This expresses an interesting relationship between the work distributions of a qubit within a complexity class of algorithms. Whilst we investigate the manifestation of these fluctuations in the work distribution associated to the trace estimation algorithm in the following section, to understand the more subtle physics of these fluctuations and their impact further studies into the explicit stochastic Lindbladian dynamics of this qubit would need to be carried out.

\section{The thermodynamics of \textbf{DQC1} trace estimation}
\label{sec:dqc1_thermo}
Once more focusing on computational inputs with only real trace contributions, the thermodynamics of the trace estimation algorithm is straightforward. In trace estimation, the logical qubit being used as a control qubit, is out of equilibrium with the maximally mixed ancillary register (which corresponds to an infinite temperature state). The more polarised the logical qubit, as quantified by $\alpha$, the more out of equilibrium it is with the ancillary register. The algorithm allows for these two quantum systems to interact and begin to equilibriate, with different trace estimations leading to different degrees of equilibration. To analyse the thermodynamics at play we can use the generalised Landauer equality or non-equilibrium 2nd Law~\cite{esposito_eq,landauer_2014}
\begin{gather}
    \langle \Sigma \rangle = \beta \langle Q \rangle + \Delta S = \mathcal{I}(S':R') + D(\rho'_R||\rho_R)
\end{gather}
as presented in Appendix~\ref{sec:energetic_laws} where the relationship
\begin{align}
    \Delta E_C &= -\langle Q_A \rangle  = \langle W_C \rangle
\intertext{is derived and the subscripts $C$ and $A$ denote the logical qubit and ancillary register respectively, giving}
 \langle W_C \rangle &= \text{tr}\left\{H_C(\rho'_C - \rho_C)\right\} \nonumber\\
 &= \alpha \omega \left(1 - \frac{\text{Re}\{\text{tr}U\}}{2^n}\right).
 \label{work_control}
\end{align}
 We see that the heat exchanged with the logical qubit is directly dependent on what trace is being estimated where the farther the trace being estimated is from that of the identity, the more the energy changes. If the trace being estimated is that of the identity, the Hamiltonian of the ancillary register will not be changed by an external agent in the implementation of this unitary. As a result, no heat is exchanged with the control qubit and so its energy will not change. On the other hand, the trace estimation of some generic unitary necessitates an external agent changing the Hamiltonian of the ancillary qubits, doing work on these qubits. This change in Hamiltonian can lead to interactions between the logical and ancillary qubits in trace estimation, and so a change in its energy.

The pre-factor $\alpha$ acts as a \textit{temperature gradient}, indicating that the greater the temperature difference at the outset, the greater the possible energy change through equilibration. The degree of equilibration is then being constrained by the computational input as the system-bath interaction given by the controlled-unitary clearly impacts the thermodynamics experienced by the register.

\begin{figure}[t]
    \centering
    \includegraphics[width=0.45\textwidth]{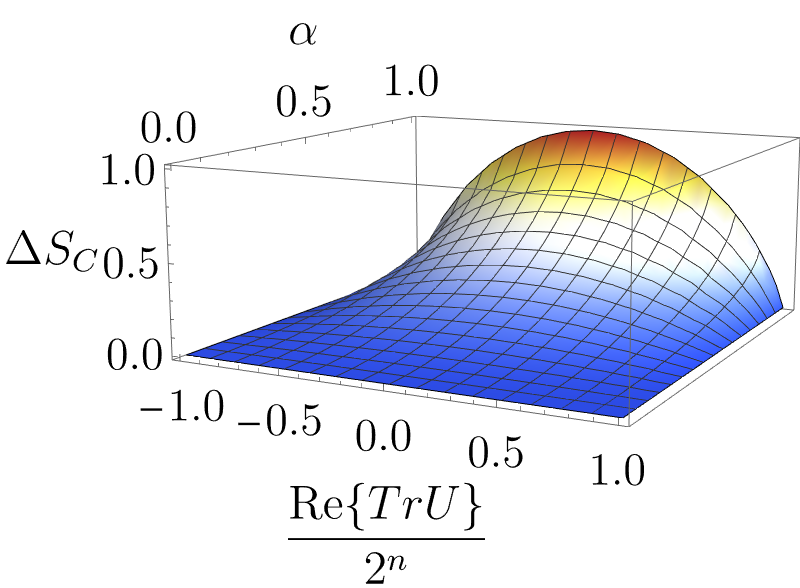}
    \caption{Change in entropy of the first qubit in the  ($\Delta S_C = S(\rho'_C) - S(\rho_C))$, for different normalised trace contributions $\left(\frac{\text{Re}\{\text{tr}U\}}{2^n}\right)$ and values of polarisation $\alpha$.}
    \label{fig:entropy}
\end{figure}

Turning our attention to the von Neumann entropy $S(\rho) = - \text{tr}\{\rho\ln \rho\}$, the entropy of the initial and final state before measurement are
\begin{align}
S(\rho_C) &= H_2\left(\frac{1-\alpha}{2}\right) \nonumber\\
S(\rho'_C) &= H_2\left(\frac{1-\alpha\frac{\text{Re}\{\text{tr}U\}}{2^n}}{2}\right) \nonumber
\end{align}
which are both binary entropy functions $H_2(x) = -  (1-x)\ln(1-x) - x\ln(x)$. The change in entropy, $\Delta S_C$, is shown in Figure~\ref{fig:entropy}. On examination, we see that the entropy is in general increasing satisfying the 2nd law of thermodynamics and that there is dependence on both the normalised trace being estimated and the polarisation of the logical qubit. As discussed in Section~\ref{sec:Crooks}, the dynamics of the control qubit respects the Tasaki-Crooks relation, where the entropy production has contributions only from the change in entropy of this qubit. From Figure~\ref{fig:entropy} we see that the change in entropy has dependence on the trace being estimated but that it is also significantly impacted by the temperature of the logical qubit through $\alpha$. As such, the hotter the logical qubit, the less entropy production and the more likely thermal fluctuations become. We can indirectly corroborate this insight by looking at the work statistics of the logical qubit, examining its nonequilibrium thermodynamics.

\begin{figure*}[!t]
	\subfloat[$\alpha = 1$]{\includegraphics[width=2in]{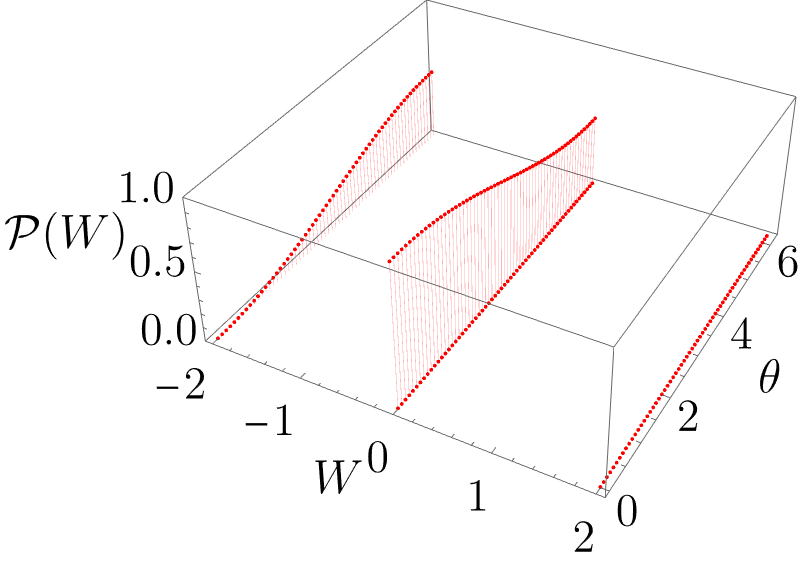}}
	\hfill
	\subfloat[$\alpha = 0.5$]{\includegraphics[width=2in]{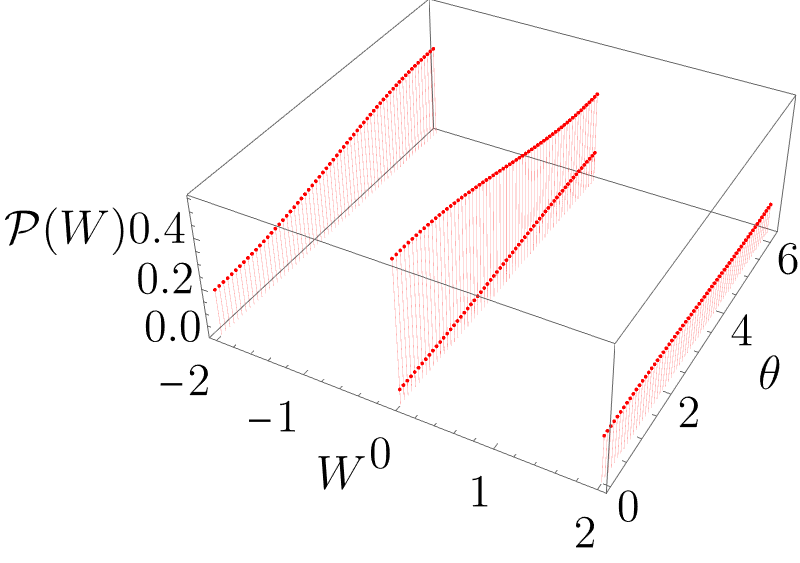}}
	\hfill
	\subfloat[$\alpha = 0$]{\includegraphics[width=2in]{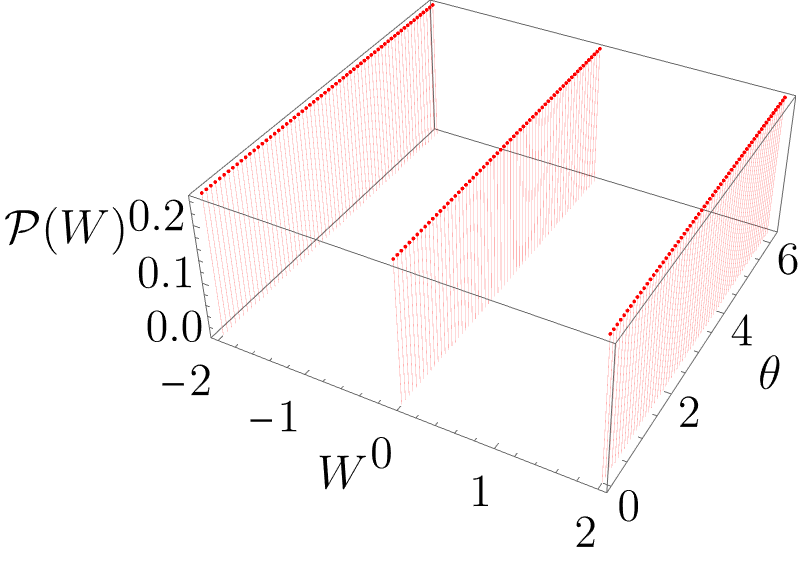}}
	\caption{Work distributions for the logical qubit in \textbf{DQC1} trace estimation with different values of $\alpha$. In these plots, $\theta$ is the parameter from Eq.~\eqref{iSWAP} giving different two qubit unitaries whose trace is estimated and $\omega = 1$. As $\alpha$ becomes smaller and so the effective temperature of the logical qubit becomes larger, fluctuations become more apparent in the work distribution.}
\label{workdistributions}
\end{figure*}

A two-point-measurement scheme~\cite{work_not_an_observable,anders_review} may be implemented to investigate the work distribution associated to the logical qubit. Here we measure in the energy eigenbasis at the start and end of the protocol allowing us to establish a probability distribution associated to the work.  Whilst delineations between heat and work are often difficult to make, in our setting this is not an issue due to Eq~ \eqref{control_qubit_1stA} meaning that all energy changes detected in the logical qubit are due to work done on it, equal to the heat it receives from the maximally mixed register.

The work probability distribution is given by 
\begin{gather}
    \mathcal{P}(W) = \sum_{n,m} p_{n}p'_{m|n} \delta\left(W - (E'_m - E_n)\right)
\end{gather}
where $p_n = \bra{n}\rho_S\ket{n}$ are the probabilities associated to the initial state, $p'_{n|m} = \left|\bra{n}\rho'_S\ket{m}\right|^2$ are the conditional probabilities associated to the final state and $E_m$ and $E_n$ are the energy eigenvalues associated to the initial Hamiltonian $H = \sum_n E_n \ketbra{E_n}{E_n}$ and the final Hamiltonian $H' = \sum_m E'_m \ketbra{E'_m}{E'_m}$. In this setting, we have the same initial and final Hamiltonian $H_C = -\omega \sigma_z$ for the logical qubit at the start and end of the algorithm, before measurement. The probabilities achieved by the scheme are readily evident from the initial state $\rho_C(\alpha)$ whose populations give the \textit{initial probabilities} and the final state $\rho'_{C}(\alpha)$ whose populations are \textit{conditional} giving a distribution with the first moment of the work distribution being
\begin{gather}
    \langle W_C \rangle = \int^{\infty}_{-\infty} W_C\mathcal{P}(W_C)\, dW_C = \Delta E_C
\end{gather}
which is equal to the change in energy of the logical qubit, recovering Eq.~\eqref{work_control}. Examining the second moment we find
\begin{align}
    \text{Var}[W_C] &= \int^{\infty}_{-\infty} W^2_C\mathcal{P}(W_C)\, dW_C\\
    &=  2\omega^2\left(\frac{\text{Re}\{\text{tr}U\}}{2^n}\alpha^2-1\right)
\end{align}
showing that fluctuations in the work distribution are mainly dependent on $\alpha$, since $\omega$ is a scaling term, corroborating what we expected from the fluctuation theorem satisfied by the logical qubit, Eq.~\eqref{crooks}.

Different trace estimations will have different work distributions, with some displaying relevant nonequilibrium behaviour on examining their higher moments if the distribution is asymmetric. To make this more tangible, we consider the trace estimation of a family of $2\times2$ matrices  given by the parameterised iSWAP
\begin{gather}
\text{iSWAP}(\theta) = 
	\begin{pmatrix} 1 & 0 & 0 & 0 \\
		0 & \cos (\theta/2) & i \sin (\theta/2) & 0 \\
		0 & i \sin (\theta/2) & \cos (\theta/2) & 0 \\
		0 & 0 & 0 & 1 \\
\end{pmatrix}
\label{iSWAP}
\end{gather}
where the trace is real therefore corresponding to an effectively thermal transformation suitable to the analysis we have made, and it is a function of $\theta$ given by $\frac{\text{Re}\{\text{tr}U\}}{4} = \cos^2(\frac{\theta}{4})$. In Figure~\ref{workdistributions} we plot the work distribution for different values of $\theta$ giving us an insight into how the nonequilibrium work varies for the trace estimation of this family of unitaries. We also plot these work distributions for different values of $\alpha$, showing how fluctuations in the work distribution become more prevalent for higher temperatures of the logical qubit, i.e. lower values of $\alpha$.

\section{Discussion \& Conclusions}
%
In this work we have treated the \textbf{DQC1} complexity class as an open quantum system whose dynamics we captured through the formalism of quantum channels. We have shown that these channels are unital and therefore respect the Tasaki-Crooks theorem. We have used this framework to investigate how the computational task of trace estimation can be related to the thermodynamics experienced by the logical qubit in the \textbf{DQC1} register, in both average and non-equilibrium settings. Restricting our analysis to unitaries with real trace, we have investigated a purely thermal process associated to changes in populations in thermal states.

In this setting we have shown that the average work done is directly related to the real part of the trace being estimated. More broadly, calculating the non-equilibrium work distribution shows that the trace estimation of different unitaries makes use of different trajectories in the evolution of the logical qubit throughout the computation.
%
%
%
We have further shown that the thermodynamics of a system carrying out a computation impacts the result of the computation itself and that that backwards processes become more likely for less entropy being produced. This last result highlights the effect of fluctuations of the system at the end of a computation, before measurement, which effectively yield a different state than one would expect to measure.

We believe that it would be of interest to investigate the thermodynamics of other complexity classes of quantum algorithms; the idea of treating the dynamics required to carry out an algorithm as a channel can be straightforwardly extended to other protocols. We note that frameworks have also been developed to investigate the quantum resources a quantum channel is capable of generating over a Hilbert space~\cite{entanglement_channels}. Applied together with the approach we describe in this work, this could open up a new route for exploring the thermodynamics of creating correlations in quantum algorithms, as has been explored in \cite{FrankEntropy} using catalytic thermal operations.

\section*{Acknowledgements}The authors acknowledge fruitful discussions with and comments from Tony J. G. Apollaro, Chris Cade, Bari\c s \c Cakmak, Sebastian Deffner, Nicolai Friis, Marcus Huber and Philip Taranto. In particular JX thanks Gabriel Landi who provided great insight into how to think about entropy production, Elisa B\"aumer whose encouragement led to this work being pursued and Mirko Consiglio for careful checking of derivations at airports. JX is supported by the Endeavour Scholarships Scheme a project which is part-financed by the European Social Fund and an IPAS+ Grant under the project \textit{MITIQ - Quantum Information and Thermodynamics in Ireland and Malta}. SC is supported by the Science Foundation Ireland Starting Investigator Research Grant ``SpeedDemon" No. 18/SIRG/5508. JG is
funded by a Science Foundation Ireland-Royal Society University Research Fellowship, the European Research Council
Starting Grant ODYSSEY (Grant Agreement No. 758403).

\providecommand{\noopsort}[1]{}\providecommand{\singleletter}[1]{#1}%

\onecolumngrid
	\appendix
\section*{Appendices}
\section{Examining Unitality}
\label{sec:Unitality}
Consider a quantum operation described by a Unitary operation $U$ on a system $\rho_S$ of dimension $d_S$ and an environment $\rho_B$ of dimension $d_B$ given as
\begin{align}
\mathcal{E}(\rho_S) &= \text{tr}_B\left\{U\rho_S\otimes\rho_B U^\dagger \right\} \\
&= \sum^{d_B}_{i,j}\bra{i} U \left( \rho_S \otimes \sum^{d_B}_{j=1} b_j\ketbra{j}{j} \right)U^\dagger \ket{i} \\
&= \sum^{d_B}_{i,j} b_j \bra{i} U \ket{j}\rho_S\bra{j}  U^\dagger \ket{i}
\end{align}
where $\ket{i},\ket{j}$ are eigenkets of the environment and $b_j$ are the eigenvalues of the initial state of the environment. The Kraus representation of this channel will then be expressed through Kraus operators of form
\begin{align}
K_{i,j} &= \sqrt{b_j}\bra{i} U \ket{j}    \end{align}
allowing us to examine the Unitality of this channel through the expression
\begin{gather}
 \sum^{d_B}_{i,j} K_{i,j}K^\dagger_{i,j} =  \sum^{d_B}_{i,j} b_j \bra{i} U \ketbra{j}{j} U^\dagger \ket{i} \label{unitality}
 \end{gather}
 which we can examine by cases.
 \subsection{Separable Unitaries}
 A separable unitary can be split over the dimensions of the system and environment as $U = U_S\otimes U_B$ so in \eqref{unitality} this gives
 \begin{align}
\sum^{d_B}_{i,j} K_{i,j}K^\dagger_{i,j} &=  \sum^{d_B}_{i,j} b_j \bra{i} U_S \otimes U_B \ketbra{j}{j} U_S^\dagger U_B^\dagger \ket{i} \\
&= \sum^{d_B}_{i,j} b_j U_S\bra{i} U_B \ketbra{j}{j} U_B^\dagger \ket{i}U_S^\dagger\\
&= \sum^{d_B}_{i,j} b_j U_S u_{i,j}u^*_{j,i} U_S^\dagger \\
&=  I_S
 \end{align}
where we have used the property that the eigenvalues of density matrices sum to 1 and that the rows and columns of unitary matrices form an orthonormal basis. Seperable unitaries in a dilation give a unital channel for any environment used in the dilation.
 \subsection{Controlled Unitaries}
 If the system of interest is a single qubit controlling a unitary operation being carried over the environment in the dilation, we can show that the resultant channel on this single qubit is unital independently of the initial state of the environment. This is sensible as such a channel on a quantum system being used to control a unitary operation on the environment being traced out is the identity channel, making it trivially unital. We present the below derivation as the the presentation is still insightful. A controlled unitary can be expressed as
 \begin{gather}
     \ketbra{0}{0}\otimes \sum^{d_B}_{k} \ketbra{k}{k} + \ketbra{1}{1} \otimes \sum^{d_B}_{l,m} u_{l,m} \ketbra{l}{m}
 \end{gather}
 and substituting in \eqref{unitality} we have
\begin{align}
&\sum^{d_B}_{i,j} K_{i,j}K^\dagger_{i,j}\\
&=\sum^{d_B}_{\substack{i,j\\k,l,m,\\n,p,q}} b_j \bra{i} \left(\ketbra{0}{0}\otimes \ketbra{k}{k} + \ketbra{1}{1} \otimes u_{l,m}\ketbra{l}{m}\right)\ketbra{j}{j} \left(\ketbra{0}{0}\otimes \ketbra{n}{n} + \ketbra{1}{1} \otimes u^*_{p,q} \ketbra{p}{q}\right) \ket{i}\\
&=\sum^{d_B}_{\substack{i,j\\k,l,m,\\n,p,q}} b_j \ketbra{0}{0}\delta_{i,k}\delta_{k,j}\delta_{j,n}\delta_{n,i} + \ketbra{1}{1}\delta_{i,l}\delta_{m,j}\delta_{j,p}\delta_{q,i}u_{l,m}u^*_{p,q}
\end{align}
summing over indices repeated indices gives
\begin{gather}
\sum^{d_B}_{\substack{l,m,\\p,q}} \ketbra{0}{0} + \ketbra{1}{1}\delta_{l,q}\delta_{m,p}u_{l,m}u^*_{p,q} = I_1
\end{gather}
finally, contracting $q$ with $l$ and $p$ with $m$ and using the orthonormality of rows and columns of unitary matrices we have unitality.
 \subsection{Arbitrary Unitaries}
 We may express the unitaries in the eigenbasis of the system and the bath in \eqref{unitality} giving
 
 \begin{align}
&=\sum^{d_B}_{i,j}\sum^{d_B d_S}_{k,l,m,n} b_j \bra{i} u_{k,l} \ketbra{k}{l}\ketbra{j}{j} u^{*}_{m,n} \ketbra{m}{n}\ket{i}
\intertext{and splitting the system and bath contributions for the system-bath eigenkets we have}
&=\sum^{d_B}_{i,j}\sum^{d_B}_{\substack{k_B,l_B,\\m_B,n_B}}
\sum^{d_S}_{\substack{k_S,l_S,\\m_S,n_S}} b_j u_{k,l}u^{*}_{m,n}\braket{i|k_B}\ketbra{k_S}{l_S}\braket{l_B|j}  \braket{j|m_B}\ketbra{m_S}{n_s}\braket{n_B|i}
\intertext{where $\alpha = \alpha_S + \alpha_B$ for $k,l,m,n$}
&=\sum^{d_B}_{i,j}\sum^{d_B}_{\substack{k_B,l_B,\\m_B,n_B}}\sum^{d_S}_{\substack{k_S,l_S,\\m_S,n_S}} b_j u_{k,l}u^{*}_{m,n} \delta_{i,k_B}\delta_{l_B,j}\delta_{j,m_B}\delta_{n_B,i}\delta_{l_S,m_S} \ketbra{k_S}{n_S}\\
&=\sum^{d_B}_{\substack{k_B,l_B,\\m_B,n_B}}\sum^{d_S}_{\substack{k_S,l_S,\\m_S,n_S}} b_{l_B} u_{k,l}u^{*}_{m,n} \delta_{l,m}\delta_{k_B,n_B}\ketbra{k_S}{n_S}\\
&=\sum^{d_B}_{\substack{k_B,l_B,\\n_B}}\sum^{d_S}_{\substack{k_S,l_S,\\n_S}} b_{l_B}u_{k,l}u^{*}_{l,n}\delta_{k_B,n_B}\ketbra{k_S}{n_S} \label{unitality_condition}
\end{align}

at which point we are unable to continue simplifying without knowing the form of the unitary. In particular, what system-bath interactions it instantiates. From this equation, we see that for a given unitary one would be able to find an environment which results in a unital channel but a general result obtained in this line of analysis seems unlikely and out of scope for this work.
\section{Effective temperature for the control qubit in the isothermal case}
\label{sec:effective_temp}
A qubit whose energy eigenbasis is also its computational basis may be described by the Hamiltonian
\begin{gather}
    H = -\omega \sigma^z
\end{gather}
and a thermal state relative to this Hamiltonian at inverse temperature $\beta$ is expressed as \begin{gather}
    \rho_\beta = \frac{e^{-\beta H}}{\text{tr}\{e^{-\beta H}\}} = \begin{pmatrix}
        \frac{1 + \tanh(\beta \omega)}{2} & 0 \\ 0 & \frac{1 - \tanh(\beta \omega)}{2}
    \end{pmatrix}.
\end{gather}
In our isothermal analysis of the trace estimation algorithm, the logical qubit is taken to have an initial Hamiltonian $H_s = -\omega \sigma^z$ which then changes to $H'_s = -\omega' \sigma^z$ as a result of work done to implement the controlled unitary gate. We stipulate for the isothermal case that the temperature of the controlled qubit remain fixed giving the thermal states
\begin{align}
    \rho_\beta = \frac{e^{-\beta H}}{\text{tr}\{e^{-\beta H_s}\}} = \begin{pmatrix}
        \frac{1 + \tanh(\beta \omega)}{2} & 0 \\ 0 & \frac{1 - \tanh(\beta \omega)}{2}
    \end{pmatrix} &&     \rho'_\beta = \frac{e^{-\beta H'_s}}{\text{tr}\{e^{-\beta H'_s}\}} = \begin{pmatrix}
        \frac{1 + \tanh(\beta \omega')}{2} & 0 \\ 0 & \frac{1 - \tanh(\beta \omega')}{2}
    \end{pmatrix}.
\end{align}
The initial and final reduced states of the logical qubit before measurement are diagonal in the energy eigenbasis 
\begin{align}
    \rho = \frac{I+\alpha \sigma^z}{2} = \begin{pmatrix}
        \frac{1 + \alpha}{2} & 0 \\ 0 & \frac{1 - \alpha}{2}
    \end{pmatrix} &&     \rho' = \frac{I+\alpha\frac{\text{Re}\{\text{tr} U\}}{2^n}\sigma^z}{2} = \begin{pmatrix}
        \frac{1 + \alpha\frac{\text{Re}\{\text{tr} U\}}{2^n}}{2} & 0 \\ 0 & \frac{1 - \alpha\frac{\text{Re}\{\text{tr} U\}}{2^n}}{2}
    \end{pmatrix}
\end{align}
and so take the form of the above thermal states leading to the relations 
\begin{align}
    \beta \omega = \arctanh(\alpha) && \beta \omega' = \arctanh\left(\alpha\frac{\text{Re}\{\text{tr} U\}}{2^n}\right)
\end{align}
where we have related the polarisation of the control qubit to the inverse temperature and a constant related to the energy gap of the qubit's basis states. In this way, we can fix $\beta$ and $\alpha$ at the outset of the process allowing us to formulate $\omega'$ 
\begin{gather}
    \omega' = \omega\left(\arctanh(\alpha)\right)^{-1}\arctanh\left(\alpha\frac{\text{Re}\{\text{tr} U\}}{2^n}\right).
\end{gather}
such that an external agent with control over this parameter could carry out work on the qubit taking it from the initial state $\rho$ to a final state $\rho'$ isothermally.

\section{Deriving the Choi Representation for the \textbf{DQC1} Trace Estimation Algorithm}
\label{Appendix:Choi}
Consider the circuit for the \textbf{DQC1} trace estimation protocol where the evolution of the state of the first qubit as a result of this protocol, before measurement, can be modelled by the quantum operation
\begin{gather}
    \mathcal{E}(\rho) = \text{tr}_n\left\{V\left(\rho\otimes \frac{I_n}{2}\right)V^\dagger\right\}.
\end{gather}
where $V = (H \otimes I_n)\left(\ketbra{0}{0}\otimes I_n + \ketbra{1}{1}\otimes U\right) (H \otimes I_n)$. By the Choi-Kraus Theorem we can associate a Kraus representation to this map
\begin{gather}
  \mathcal{E}(\rho_S) = \sum_{i}\Lambda_i \rho_S \Lambda^\dagger_i \, : \, \sum_{i} = \Lambda_i^\dagger \Lambda_i = I_1
\end{gather}
where $\Lambda_i$ are the associated Kraus operators. These Kraus operators may be derived from the unitary $V$ in the Stinespring dilation as \begin{gather}
    \Lambda_{i,j} = \bra{i_B}V\ket{\widetilde{j_B}} = \sqrt{b_j}\bra{i_B}V\ket{j_B}
\end{gather}
where $\ket{i_B}$ is an eigenket of the environment, $\ket{\tilde{j}}$ is a eigenket for the inital state of the environment and $\ket{b_i}$ is an eigenvalue of the initial state of the environment. The Kraus operators are not unique and dependent on the choice of basis for the environment. For the case of 2+1 \textbf{DQC1} with the general two qubit unitary
\begin{gather}
U = \sum^{4}_{i,j=1} u_{ij} \ketbra{i}{j} = \left(
\begin{array}{cccc}
 u_{11} & u_{12} & u_{13} & u_{14} \\
 u_{21} & u_{22} & u_{23} & u_{24} \\
 u_{31} & u_{32} & u_{33} & u_{34} \\
 u_{41} & u_{42} & u_{43} & u_{44} \\
\end{array}
\right) \, : \, u_{ij} \in \mathbb{C}
\end{gather}
whose trace we are estimating, this gives the 16 Kraus operators using the computational basis.
\begin{center}
\begin{tabular}{ccc}
 $K_{00} = \left(
\begin{array}{cc}
 \frac{1}{4} \left(u_{11}+1\right) & \frac{1}{4} \left(1-u_{11}\right) \\
 \frac{1}{4} \left(1-u_{11}\right) & \frac{1}{4} \left(u_{11}+1\right) \\
\end{array}
\right)$
&
$K_{01} = \left(
\begin{array}{cc}
 \frac{u_{13}}{4} & -\frac{u_{13}}{4} \\
 -\frac{u_{13}}{4} & \frac{u_{13}}{4} \\
\end{array}
\right)$
& 
$K_{02} = \left(
\begin{array}{cc}
 \frac{u_{12}}{4} & -\frac{u_{12}}{4} \\
 -\frac{u_{12}}{4} & \frac{u_{12}}{4} \\
\end{array}
\right)$
\\  
$K_{03} = \left(
\begin{array}{cc}
 \frac{u_{14}}{4} & -\frac{u_{14}}{4} \\
 -\frac{u_{14}}{4} & \frac{u_{14}}{4} \\
\end{array}
\right)$
&
$K_{10} =  \left(
\begin{array}{cc}
 \frac{u_{31}}{4} & -\frac{u_{31}}{4} \\
 -\frac{u_{31}}{4} & \frac{u_{31}}{4} \\
\end{array}
\right)$
& 
$K_{11} =  \left(
\begin{array}{cc}
 \frac{1}{4} \left(u_{33}+1\right) & \frac{1}{4} \left(1-u_{33}\right) \\
 \frac{1}{4} \left(1-u_{33}\right) & \frac{1}{4} \left(u_{33}+1\right)\\
\end{array}
\right)$
\\
$K_{12} = \left(
\begin{array}{cc}
 \frac{u_{32}}{4} & -\frac{u_{32}}{4} \\
 -\frac{u_{32}}{4} & \frac{u_{32}}{4} \\
\end{array}
\right)$
& 
$K_{13} = \left(
\begin{array}{cc}
 \frac{u_{34}}{4} & -\frac{u_{34}}{4} \\
 -\frac{u_{34}}{4} & \frac{u_{34}}{4} \\
\end{array}
\right)$ 
&
$K_{20} = \left(
\begin{array}{cc}
 \frac{u_{21}}{4} & -\frac{u_{21}}{4} \\
 -\frac{u_{21}}{4} & \frac{u_{21}}{4} \\
\end{array}
\right)$\\
$
K_{21} = \left(
\begin{array}{cc}
 \frac{u_{23}}{4} & -\frac{u_{23}}{4} \\
 -\frac{u_{23}}{4} & \frac{u_{23}}{4} \\
\end{array}
\right)$ 
& 
$
K_{22} = \left(
\begin{array}{cc}
 \frac{1}{4} \left(u_{22}+1\right) & \frac{1}{4} \left(1-u_{22}\right) \\
 \frac{1}{4} \left(1-u_{22}\right) & \frac{1}{4} \left(u_{22}+1\right) \\
\end{array}
\right)$ 
&
$
K_{23} = \left(
\begin{array}{cc}
 \frac{u_{24}}{4} & -\frac{u_{24}}{4} \\
 -\frac{u_{24}}{4} & \frac{u_{24}}{4} \\
\end{array}
\right)$\\
$
K_{30} = \left(
\begin{array}{cc}
 \frac{u_{41}}{4} & -\frac{u_{41}}{4} \\
 -\frac{u_{41}}{4} & \frac{u_{41}}{4} \\
\end{array}
\right)$
& 
$
K_{31} = \left(
\begin{array}{cc}
 \frac{u_{43}}{4} & -\frac{u_{43}}{4} \\
 -\frac{u_{43}}{4} & \frac{u_{43}}{4} \\
\end{array}
\right)$ 
&
$
K_{32} = \left(
\begin{array}{cc}
 \frac{u_{42}}{4} & -\frac{u_{42}}{4} \\
 -\frac{u_{42}}{4} & \frac{u_{42}}{4} \\
\end{array}
\right)$
\\
&
$
K_{33} = \left(
\begin{array}{cc}
 \frac{1}{4} \left(u_{44}+1\right) & \frac{1}{4} \left(1-u_{44}\right) \\
 \frac{1}{4} \left(1-u_{44}\right) & \frac{1}{4} \left(u_{44}+1\right) \\
\end{array}
\right)$
&
\end{tabular}
\end{center}
Using the Kraus representation for $\mathcal{E}(\rho)$ provided by these operators we may construct the Choi state corresponding to this channel by the Choi-Jamio\l{}kowski isomorphism
\cite{modi_operational_dynamics,choi,jamiolkowski}
\begin{gather}
\Upsilon_\mathcal{E} = \left(\mathcal{E}(\rho) \otimes I_{1}\right)\ketbra{\varphi}{\varphi}
\end{gather}
allowing us to obtain
\begin{gather}
\Upsilon_\mathcal{E} = \frac{1}{16}\begin{pmatrix}
4 + \sum^{4}_{i,j} |u_{ij}|^2+\text{Re}\{\text{tr}U\} & -\sum^{4}_{i,j} |u_{ij}|^2+ i\text{Im}\{\text{tr}U\} & -\sum^{4}_{i,j} |u_{ij}|^2+i\text{Im}\{\text{tr}U\} & 4+ \sum^{4}_{i,j} |u_{ij}|^2 + \text{Re}\{\text{tr}U\}\\
-\sum^{4}_{i,j} |u_{ij}|^2 -i\text{Im}\{\text{tr}U\} & 4 + \sum^{4}_{i,j} |u_{ij}|^2  - \text{Re}\{\text{tr}U\} & 4 + \sum^{4}_{i,j} |u_{ij}|^2  - \text{Re}\{\text{tr}U\} & - \sum^{4}_{i,j} |u_{ij}|^2 -i\text{Im}\{\text{tr}U\}\\
- \sum^{4}_{i,j} |u_{ij}|^2 -i\text{Im}\{\text{tr}U\} & 4 + \sum^{4}_{i,j} |u_{ij}|^2 - \text{Re}\{\text{tr}U\} & 4 + \sum^{4}_{i,j} |u_{ij}|^2  - \text{Re}\{\text{tr}U\} & - \sum^{4}_{i,j} |u_{ij}|^2-i\text{Im}\{\text{tr}U\}\\
4 +\sum^{4}_{i,j} |u_{ij}|^2 +  \text{Re}\{\text{tr}U\} & - \sum^{4}_{i,j} |u_{ij}|^2 i\text{Im}\{\text{tr}U\} & - \sum^{4}_{i,j} |u_{ij}|^2+ i\text{Im}\{\text{tr}U\} & 4 + \sum^{4}_{i,j} |u_{ij}|^2 + \text{Re}\{\text{tr}U\}
\end{pmatrix} 
\end{gather}
which can clearly scale to $n+1$ \textbf{DQC1} trace estimation as follows
\begin{gather}
 \hspace*{-1cm} 
\Upsilon_\mathcal{E} = \frac{1}{2^{n+2}}\begin{pmatrix}
2^n + \sum^{2^n}_{i,j} |u_{ij}|^2+\text{Re}\{\text{tr}U\} & -\sum^{2^n}_{i,j} |u_{ij}|^2+ i\text{Im}\{\text{tr}U\} & -\sum^{2^n}_{i,j} |u_{ij}|^2+i\text{Im}\{\text{tr}U\} & 2^n+ \sum^{2^n}_{i,j} |u_{ij}|^2 + \text{Re}\{\text{tr}U\}\\
-\sum^{2^n}_{i,j} |u_{ij}|^2 -i\text{Im}\{\text{tr}U\} & 2^n + \sum^{2^n}_{i,j} |u_{ij}|^2  - \text{Re}\{\text{tr}U\} & 2^n + \sum^{2^n}_{i,j} |u_{ij}|^2  - \text{Re}\{\text{tr}U\} & - \sum^{2^n}_{i,j} |u_{ij}|^2 -i\text{Im}\{\text{tr}U\}\\
- \sum^{2^n}_{i,j} |u_{ij}|^2 -i\text{Im}\{\text{tr}U\} & 2^n + \sum^{2^n}_{i,j} |u_{ij}|^2 - \text{Re}\{\text{tr}U\} & 2^n + \sum^{2^n}_{i,j} |u_{ij}|^2  - \text{Re}\{\text{tr}U\} & - \sum^{2^n}_{i,j} |u_{ij}|^2-i\text{Im}\{\text{tr}U\}\\
2^n +\sum^{2^n}_{i,j} |u_{ij}|^2 +  \text{Re}\{\text{tr}U\} & - \sum^{2^n}_{i,j} |u_{ij}|^2 -i\text{Im}\{\text{tr}U\} & - \sum^{2^n}_{i,j} |u_{ij}|^2+ i\text{Im}\{\text{tr}U\} & 2^n + \sum^{2^n}_{i,j} |u_{ij}|^2 + \text{Re}\{\text{tr}U\}
\end{pmatrix}.
\end{gather}
Finally, this Choi representation is only dependent on the trace contributions since for unitary matrices we have that the square of the Frobenius norm 
\begin{gather}
   {||U||_F}^2 = \sum^{2^n}_{i,j=1} |u_{ij}|^2 = \text{tr}\{U^{\dagger}U\} = \text{tr}\{I_n\} = 2^n
\end{gather}
giving the Choi state as in Eq.(6)
\begin{gather}
\Upsilon_\mathcal{E} = \begin{pmatrix}
\frac{1 + \frac{\text{Re}\{\text{tr}U\}}{2^n}}{2} & i\frac{\text{Im}\{\text{tr}U\}}{2^{n+1}} & i\frac{\text{Im}\{\text{tr}U\}}{2^{n+1}} & \frac{1 + \frac{\text{Re}\{\text{tr}U\}}{2^n}}{2}\\
-i\frac{\text{Im}\{\text{tr}U\}}{2^{n+1}} & \frac{1 - \frac{\text{Re}\{\text{tr}U\}}{2^n}}{2} & \frac{1 - \frac{\text{Re}\{\text{tr}U\}}{2^n}}{2} &-i\frac{\text{Im}\{\text{tr}U\}}{2^{n+1}}\\
 -i\frac{\text{Im}\{\text{tr}U\}}{2^{n+1}} & \frac{1 - \frac{\text{Re}\{\text{tr}U\}}{2^n}}{2} & \frac{1 - \frac{\text{Re}\{\text{tr}U\}}{2^n}}{2} &-i\frac{\text{Im}\{\text{tr}U\}}{2^{n+1}}\\
\frac{1 + \frac{\text{Re}\{\text{tr}U\}}{2^n}}{2} & i\frac{\text{Im}\{\text{tr}U\}}{2^{n+1}} &  i\frac{\text{Im}\{\text{tr}U\}}{2^{n+1}} & \frac{1 + \frac{\text{Re}\{\text{tr}U\}}{2^n}}{2}
\end{pmatrix}\\
\Upsilon_\mathcal{E}=
\frac{1}{2}\left[
\begin{array}{c|c}
I_1 + \frac{\text{Re}\{\text{tr}U\}}{2^n}\sigma_z + \frac{\text{Im}\{\text{tr}U\}}{2^n}\sigma_y  & \sigma_x + i\frac{\text{Re}\{\text{tr}U\}}{2^n}\sigma_y + i\frac{\text{Im}\{\text{tr}U\}}{2^n}\sigma_z \\
\hline
\sigma_x - i\frac{\text{Re}\{\text{tr}U\}}{2^n}\sigma_y - i\frac{\text{Im}\{\text{tr}U\}}{2^n}\sigma_z & I_1 - \frac{\text{Re}\{\text{tr}U\}}{2^n}\sigma_z - \frac{\text{Im}\{\text{tr}U\}}{2^n}\sigma_y
\end{array}
\right]
\end{gather}
On examination, one finds that this is a rank-2 Choi Matrix meaning that this map can be described by two unwieldy Kraus Operators.

\section{Two Energetic 1st Laws for \textbf{DQC1}}
\label{sec:energetic_laws}
The main feature of \textbf{DQC1} algorithms is that they make use of a register
\begin{gather}
\frac{I_1 +\alpha \sigma^z_1}{2} \otimes \frac{I_n}{2^n}
\end{gather}
where $\alpha = 1$ in the original \textit{clean} \textbf{DQC1} case~\cite{DQC1} and $\alpha \in [0,1]$ in the \textit{non-clean} case~\cite{datta_2005,non_clean}, and it is only the first qubit which is measured. In this sense, the first qubit is the \textit{logical} or \textit{special} qubit and the qubits in the maximally mixed state may be thought of as ancillas. The letter $C$ is used here as in the trace estimation setting the logical qubit acts as a control qubit. Let us label the first qubit $\rho_C = \frac{I_1 +\alpha \sigma^z_1}{2}$ and the mixed qubits $\rho_A = \frac{I_n}{2^n}$.

The equation
\begin{gather}
    \langle \Sigma \rangle = \beta \langle Q \rangle + \Delta S = \mathcal{I}(S':R') + D(\rho'_R||\rho_R)
\end{gather}
expresses the average entropy production in terms of the correlations formed across a system and a reservoir, quantified by the mutual information, and relative entropy of the states of the reservoir at the start and end of a nonequilibrium process where initially, the reservoir is in a thermal state and the system and reservoir are separable. This expression was originally explored in~\cite{esposito_eq} to split the entropy production into a thermodynamically reversible contribution due to heat exchange and thermodynamically irreversible contribution due to the formation of correlations but was later reformulated~\cite{landauer_2014} as presented above which can be regarded as nonequilibrium 2nd Law. Clearly, the register of \textbf{DQC1} algorithms is amenable to this framework so let's start our thermodynamic analysis here.

From the perspective of the logical qubit, we can consider it to be the thermodynamic system of interest and the maximally mixed register as the reservoir, which is thermal relative to some $H_A = \sum_i \omega_i \sigma^z_i$ local Hamiltonian. Here, we have $D(\rho'_A || \rho_A) = 0$ since $\rho'_A = \rho_A = \frac{I_n}{2^n}$ and the mutual information
\begin{gather}
    \mathcal{I}\left(\rho'_C \, :\, \rho'_A  \right) = S(\rho'_C) + S(\rho'_A) - S(\rho'_{CA}).
\end{gather}
Since von Neumann entropy is invariant under the action of a unitary we find
\begin{gather}
    S(\rho'_{CA}) = S(V\rho_{C}\otimes \rho_A V^\dagger)\\
    = S(\rho_C \otimes \rho_A) = S(\rho_C)  + S(\rho_A) 
\end{gather}
where $V \in SU(N+1)$ is the global unitary representing a given \textbf{DQC1} algorithm and the additivity of the von Neumann entropy over separable states was used giving
\begin{gather}
    \mathcal{I}(\rho'_C \, : \, \rho'_A) = S(\rho'_C) - S(\rho_C) = \Delta S_C 
\end{gather}
which implies that the entropy production of the control qubit is equal to its change in entropy and it transfers no heat to the maximally mixed register. 
\begin{align}
    \langle \Sigma_C \rangle = \Delta S_C && \beta_A\langle Q_C \rangle = 0
\label{ent_prod_control}
\end{align}
This can also be seen from the fact that the ancillary component of the register remains in the maximally mixed state throughout the computation giving
\begin{gather}
    \beta_A \langle Q_C \rangle = \text{tr}\left\{(\rho'_A - \rho_A )H_A\right\} = 0
\end{gather}
and the 1st law for the logical qubit 
\begin{gather}
    \Delta E_c = \langle W_C \rangle.
\label{control_qubit_1st}
\end{gather}
Shifting perspectives, we can now focus on the maximally mixed register as our thermodynamic system of interest and consider the logical qubit, whose initial state is thermal relative to $H_C = \omega \sigma_z$,  to be the reservoir. We begin with the entropy where we now have $\Delta S_A = 0$ since the state of the ancillary register is unchanged and all entropy production will be associated to heat exchanged. For the same reason, the 1st law in this context is \begin{gather}
\langle W_A \rangle = - \langle Q_A \rangle.
\end{gather}
The mutual information remains the same as (6) due to its symmetry but the relative entropy term is now
\begin{gather}
    D(\rho'_C||\rho_C) = - S(\rho'_C) + \text{tr}\left\{ \rho'_C\text{ln}\rho_C \right\}
\end{gather}
giving the entropy production for the ancillas
\begin{gather}
    \langle \Sigma_A \rangle = \Delta S_C - S(\rho'_C) + D(\rho'_C||\rho_C)\\
    = \text{tr}\left\{(\rho'_C - \rho_C) \text{ln} \rho_C\right\} = \Delta E_C\\
    \implies \langle \Sigma_A \rangle = \beta_C \langle Q_A \rangle = \Delta E_C
\end{gather}
and so the entropy production of the ancillary register is equal to the heat it exchanges with the logical qubit, which is equal to its energy change. In this sense, the thermodynamics of this complexity class is straightforward.

As a result of the equality form of Landauer's principle~\cite{esposito_eq,landauer_2014} the thermodynamics of any \textbf{DQC1} algorithm may be summarised as heat being exchanged from the ancillary register to the logical qubit which varies depending on how much the entropy of the first qubit changes. This will be dependent on correlations formed between partitions involving the logical qubit and parts of the ancillary register with the remaining qubits \cite{datta_2008,cade2017quantum} and the temperature difference between the qubits inherent of the \textbf{DQC1} complexity class. Whilst the choice of what is the system and what is the bath is subjective, the thermodynamics experienced in terms of the heat exchanged, is not. As is verified by looking at both these perspectives. 

It is important to appreciate that if one has \textbf{DQC1} algorithms represented by unitaries that are not energy preserving, that is 
\begin{gather}
    [V, H_S + H_C] \neq 0
\end{gather}
that the work terms in (9) and (10) may include contributions from an external agent implementing this unitary and more housekeeping would be necessary. Since the ancillary qubits are in a maximally mixed state no work done on the ancillary qubits by an external agent can be captured by this scheme. The work we see is work carried out by the ancillary qubits in exchanging this heat with the control qubit -- which is equal to the energy change in the control qubit. This cannot be said for the logical qubit in  the \textbf{DQC1} complexity class where the work carried out on this qubit can not only feature heat it acquires from the ancillary qubits through the interaction mediated by $V$ but also in changes to its local Hamiltonian which may occur during the implementation of $V$. For this reason, we dub these energetic 1st laws as opposed to standard thermodynamic 1st laws.

\end{document}